# In vivo three-dimensional brain and extremity MRI at 50 mT using a permanent magnet Halbach array

T. O'Reilly[1], W. M. Teeuwisse[1], D. de Gans[2], K. Koolstra[1], A. G. Webb[1,2]

1. C.J. Gorter Center for High Field MRI, Department of Radiology, Leiden University Medical Center, Leiden, The Netherlands.
2. Technical University Delft, The Netherlands.

Corresponding author:
Andrew Webb, C.J.Gorter Center for High Field MRI
Department of Radiology, Leiden University Medical Centre,
Albinusdreef 2, 2333 ZA Leiden. The Netherlands
Email: a.webb@lumc.nl

**Abstract**

**Purpose:** To design a low-cost, portable permanent magnet-based MRI system capable of obtaining in vivo MR images within a reasonable scan time.

**Methods:** A discretized Halbach permanent magnet array with a clear bore diameter of 27 cm was designed for operation at 50 mT. Custom built gradient coils, radiofrequency coil, gradient amplifiers and radiofrequency amplifier were integrated and tested on both phantoms and in vivo.

**Results:** Phantom results showed that the gradient non-linearity in the y- and z-directions was less than 5% over a 15 cm field-of-view and did not need correcting. For the x-direction, it was significantly greater, but could be partially corrected in post-processing. Three dimensional In vivo scans of the brain of a healthy volunteer using a turbo-spin echo sequence were acquired at a spatial resolution of 4×4×4 mm in a time of ~2 mins. $T_1$-weighted and $T_2$-weighted scans showed a good degree of tissue contrast. In addition, in vivo scans of the knee of a healthy volunteer were acquired at a spatial resolution of ~3x2x2 mm within a twelve minutes to show the applicability of the system to extremity imaging.

**Conclusion:** This work has shown that it is possible to construct a low-field MRI unit with hardware components costing less than 10000 euros, which is able to acquire human images in vivo within a reasonable data acquisition time. The system has a high degree of portability with magnet weight ~75 kg, gradient and RF amplifiers each 15 kg, gradient coils 10 kg and spectrometer 5 kg.

**Introduction**

In the past few years there has been renewed interest in designing human MRI systems with field strengths significantly below those (1.5 and 3 Tesla) used for standard clinical examinations [1]. Much of the rationale arises from the very high purchase, maintenance and siting costs associated with such systems, all of which put them out of the financial reach of much of the world's population. In addition to financial considerations, there are intrinsic advantages in terms of fewer patient contraindications and reduced specific absorption ratio (SAR) compared to higher field systems.

Although there is a significant loss in signal-to-noise ratio (SNR) at low fields, the work by the Rosen group, in particular, has shown that diagnostically-useful image quality can be achieved using fields as low as 6.5 mT by using highly efficient sequences in combination with a very homogeneous magnetic field created by a large electromagnet [2-4]. Representing a slight increase in hardware complexity, combinations of pre-polarizing electromagnets and lower field either electro- or permanent magnet based readout systems have been used for imaging human extremities [5, 6]. More sophisticated field-cycling systems [7-11] have also been used to obtain data over a wide range of field strengths, most recently in vivo data at 50 $\mu$T.

A fourth approach to low field MRI, which has the advantages of the relative simplicity of a single fixed field strength magnet, no requirement for a magnet power supply, and increased portability, is to use/design a permanent magnet. Although the original commercial C-arm or H-arm systems were large and heavy (>1000 kg) with field strengths of between 0.1 and 0.35 Tesla [12], more recently, a relatively inexpensive commercial system with significantly reduced size operating at 60 mT has shown very high quality images of the brain, with potential applications in the emergency room (www.hyperfine.io). However, even this system weights approximately 500 kg, which limits its transportability.

In order to reduce the weight of the magnet, a large number of distributed small magnets, rather than two very large disc magnets as in most permanent magnet designs, can be used. The most common geometry is a discretized version of a dipolar Halbach magnet [13], which is an optimal design in terms of producing the highest homogeneous magnet field within the structure and minimal field outside. This discrete geometry is in practise constructed from many small individual magnets. There are a number of papers in the literature which describe the design and construction of Halbach arrays for different applications, and analysis techniques for calculating both magnetic fields and inter-element forces produced by such arrays [14-16]. For NMR and MRI applications the Halbach array was first demonstrated, using the k=1 dipolar mode which produces a relatively homogeneous transverse magnetic field, in a series of papers from Blümler and associates [17-19], who constructed magnets capable of measurements on small animals and plants. A number of other small scale systems have been constructed using modifications of this basic approach [20-22]. Magnets with an in-built linear gradient have also been proposed [23] and designed [24,25], in which rotation of the magnet around the sample enables two-dimensional spatial information to be encoded and reconstructed.

Recently, our group described the design, construction and characterization of a larger 27 cm clear-bore Halbach array magnet operating at 50 mT [26]. A genetic algorithm was used to optimize the geometry of the 23-ring magnet, and the inclusion of additional shimming rings to compensate for material and construction imperfections resulted in a homogeneity of approximately 2400 parts-per-million (ppm) over a 20 cm diameter-of-spherical-volume (DSV). Using simple gradient coils, three dimensional spin-echo images at 3.5 × 3.5 × 3.5 mm spatial resolution were acquired in an imaging time of ~35 minutes from a phantom. However, this time is much too long for a useful clinical scan, and in addition the gradient coils had too high a resistance to be able to implement efficient imaging sequences with a high duty cycle. Since our major aim for a portable MRI system is to image pediatric hydrocephalus in resource-limited locations, we ultimately need the system to acquire images distinguishing between cerebrospinal fluid (CSF) and white matter (WM) and gray matter (GM). The role of imaging in pediatric hydrocephalus has recently been reviewed [27]. As decribed in this review, and earlier papers [28], 3D $T_1$-weighted and $T_2$-weighted images form the essential anatomical sequences, with other sequences supplementing these based upon the particular feature of the disease being studied. Although much higher spatial resolution can be achieved at high fields, the loss in SNR at low fields can be partially retrieved by the use of long echo trains in turbo spin echo (TSE) sequences due to the long $T_2$ value of CSF and the very low SAR intrinsic to low field MRI.

In this paper we design a long echo train turbo spin echo sequence to image the CSF in healthy adult brain. The Cartesian k-space coverage is restricted to an elliptical coverage to reduce the data acquisition time to less than three minutes for a 4 × 4 × 4 mm isotropic spatial resolution. We also show applications for extremity imaging in which higher resolution (~3 × 2 × 2 mm) in vivo images of the human knee in just over 10 minutes. Issues of temperature drift and image reconstruction in the presence of gradient non-linearities are also discussed.

**Method**

*Magnet design*

Details of the magnet design have been published previously [26] and so are minimally described here. The Halbach magnet consists of twenty three rings, with two layers of N48 neodymium boron iron (NdBFe) magnets (12 x 12 x 12 mm$^3$) per ring. The array has a clear bore of 27 cm, and a length of 50 cm between the two outer rings. The ring diameters were designed via an optimization scheme to give the highest $B_0$ homogeneity. After measuring the field using a 3D-robotic device, additional shimming was implemented by defining a 32.5 cm diameter, 28 cm long cylindrical grid of 15 rings of 60 potential magnet positions per ring, the filling of this grid was subsequently optimized using a genetic algorithm, which has previously been shown to be suitable for creating tailored magnetic fields using permanent magnets [24], that had 3 options for each magnet position; no magnet, a magnet following the k=1 Halbach orientation and finally the magnet flipped 180 degrees from that orientation. The genetic algorithm minimizes the field variation over a 20 cm diameter spherical volume with 3 × 3 × 3 mm N45 NdBFe magnets being used to fill the grid. The optimal solution occupied approximately 600 out of the 900 grid positions and once constructed reduced the field inhomogeneity from 13000 ppm to 2400 ppm with the majority of the $B_0$ deviations located towards the edges of this DSV along the z axis. The total weight of the magnet including all components is ~75 kg.

To minimize the environmental noise the entire setup was placed inside a 62.5 × 62.5 × 85 cm Faraday cage constructed from 2 mm thick aluminium sheets and 32 × 32 mm$^2$ aluminium extrusion profiles. Since the body can act as a very effective antenna, coupling to power lines in the surrounding walls for example [21], the torso extending out of the Faraday shield was wrapped in a conductive fabric (Holland Shielding Systems, Dordrecht, the Netherlands). The sheet was connected to the same ground as the Faraday cage, as shown in Figure 1.

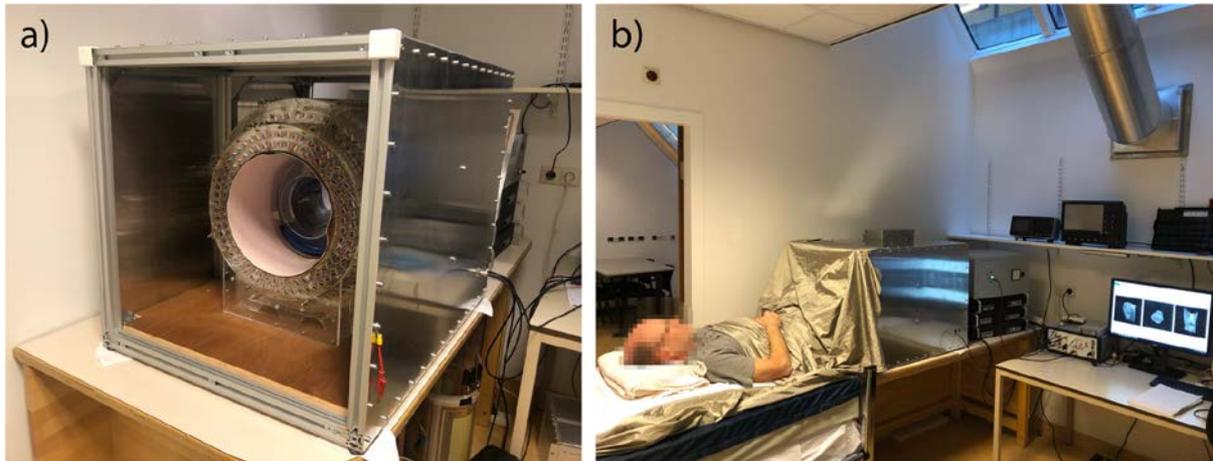

*Figure 1. A) Photograph of the magnet setup. The shims and gradients are integrated in to the bore of the magnet. The entire setup is placed inside an aluminium Faraday cage to reduce environmental noise in the setup. B) A conductive cloth is placed over the subject to reduce the noise coupled in to the system through the body.*

*Radiofrequency coil*

For the phantom and knee imaging experiments a 15 cm long, 15 cm diameter solenoid with 57 windings was used as a transmit/receive coil. The solenoid was segmented using three non-magnetic capacitors (47 pF, American Technical Ceramics E-series) spaced evenly along the length of the solenoid to avoid self-resonance effects. Impedance matching used an L-network with 33 pF and 136 (2 x 68) pF capacitors in the parallel and series arms, respectively. There was a 4 kHz shift downwards in the resonant frequency when the knee was inserted. The Q value decreased from 54 to 47 when the coil was loaded with the knee, indicating that coil loss is dominant as expected [29]. A thin copper sheet (50 µm) was placed inside the gradient coils to reduce the interaction between the self-resonant modes of the gradient coils and the RF coil as well as to reduce noise coupled in to the system from the gradient amplifier. Six 800 pF chip capacitors were placed along the seam of the sheet to reduce eddy currents induced by switching of the gradients. The resonance frequency of the RF coil increased 95 kHz when the coil was inserted in to the magnet.

For in vivo brain imaging an elliptical solenoid was constructed with a width of 18.5 cm, height of 24.5 cm and length of 20 cm with 40 windings of 1 mm diameter copper wire. It is segmented in 3 places with 82pf capacitors with a 1-30pF variable capacitor placed in parallel with one of the segmentation capacitors. Impedance matching used an L-shaped matching network with 101 pF (68 pF + 33 pF) and 150 pF (82 pF + 68 pF) capacitors in the parallel and series arms, respectively. The resonance frequency increased 233 kHz when the coil was inserted in to the magnet. The unloaded Q was 88, loading the coil with a human head resulted in a 4 kHz decrease in the resonance frequency of the coil and a decreased Q of 78.

When either coil was used without the conductive sheet in place, the noise level increased by approximately a factor-of-ten due to the "body antenna" effect. When the sheet was carefully wrapped around the entire body, and physical contact with the Faraday shield was present at several points, the noise level was essentially unchanged from a coil loaded with a tissue-mimicking phantom within the solid Faraday enclosure.

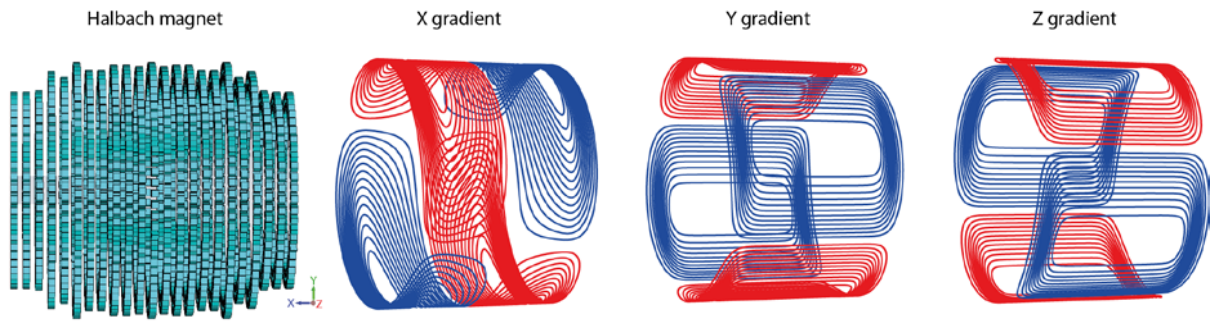

*Figure 2. Three imaging gradients are designed using the Target Field Method, adapted for the Halbach configuration with the $B_0$ field oriented across the magnet bore (left schematic): the line colour indicates the direction of current flow. The gradients were constructed using copper wires placed into a 3 mm-thick 3D printed structure.*

*Three-axis gradient coils*

In a previous publication [26] simple quadrupolar Y and Z gradients were constructed. These had high inductance 203 µH and very high resistance 10.0 Ω. Here, new gradient coils with improved performance were designed, constructed and integrated. Gradient coils were designed using the target field approach [30] adapted to produce their encoding fields commensurate with the transverse $B_0$ direction of the Halbach array. This results in y and z gradient coils which are similar in design to a quadrupolar design, but an x gradient coil with significantly improved performance compared to a simple saddle design: full mathematical details are given in de Vos et al. [31].

Gradients coils were constructed using 1.5 mm diameter copper wire pressed into 3D printed formers. The efficiencies of the x, y and z gradient coils were 0.59, 0.95 and 1.02 mT/m/A, respectively. The inductance of the x gradient was 180 µH with a resistance of 0.37 Ω, the inductances of the y and z gradients were both 225 µH with resistances of 0.4 Ω. Schematics of the wire layouts for the three gradient coils are given in Figure 2, 3D models for the X, Y and Z gradient coils are given in supporting information figure 2, 3 and 4, respectively.

*Gradient amplifiers*

Gradient power amplifiers (GPAs) on most clinical 1.5 and 3T superconducting whole body scanners have maximum gradient strengths in the range of 40-80 mT/m, with slew rates of 150-200 T/m/s. GPAs for these systems are capable of simultaneously generating a driving current of 1 kA and voltage of 2 kV over a bandwidth of tens of kHz by cascading the outputs of several PWM stages. Typically the current signal is controlled by active feedback using high-precision current sensors built into the circuit. Gain accuracy and linearity are typically below 0.05% and total harmonic distortion (THD) below 0.25% [32]. GPAs for preclinical imaging are typically four-quadrant linear designs with maximum output currents on the order of 40-80 amps, with maximum voltages 100-200 volts. These GPAs can also be used for MR microscopy, in one example Seeber et al. described a pulsed current gradient power supply for very high spatial resolution microimaging [33].

However, for lower field systems aimed at creating low-cost, low-maintenance MRI systems, such high gradient performance is not required. In this type of system, spin-echo based sequences with echo times on the order of tens of milleseconds mean that very rapid switching and rise times (high slew rates) are not required. Spatial resolutions on the order of several mm, mean that frequency encoding gradient strengths on the order of tens of milletesla per metre are sufficient and much more inexpensive linear GPA designs can be used. Wright et al. developed a desktop MRI system capable of imaging objects approximately 2 cm in diameter. The GPAs used LM12 high power op-amps capable of driving +/- 10 amps into gradient coils with an efficiency of 0.45 G/cm/A [34]. One recent example is the system constructed to run educational MRI devices: the OpenMRI project uses two OPA 549 power op-amps in a bridged configuration, with a maximum output current of 8 amps [35].

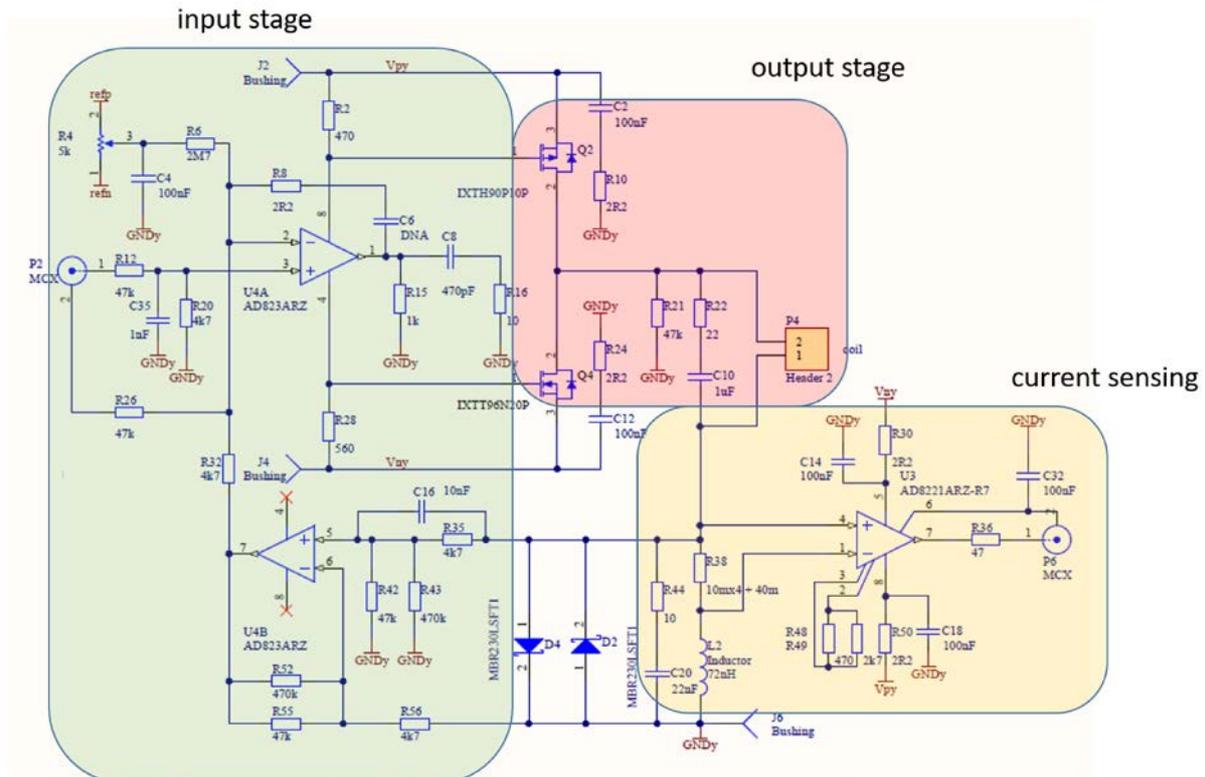

*Figure 3. Schematic of the three modules used in each of the three gradient amplifiers channels.*

Since we do not need very high currents to achieve the desired spatial resolution (~3 mm) high currents are not necessary, and the maximum voltage (15 volts) is chosen such that simple air-fan and heatsink cooling can be used. Each amplifier was constructed using a push-pull configuration. The design for each channel is shown in Figure 3. The basic design is a pair of bridged operational amplifiers (U4A and U4B), which provide voltage gain to the input of a rail-to-rail power MOSFET amplifiers which produce a current output which is independent of the load inductance. For simplicity the circuit is shown as three separate, but connected, modules, each of which is described in detail below.

The input stage consists of an operational amplifier, which provides the necessary voltage gain sufficient to drive the MOSFETS. The output stage consists of a MOSFET rail-to-rail CS stage. It consists of two power MOSFETs arranged in a push-pull configuration. The IXTT90P10P power MOSFET contains a fast intrinsic diode with VDSS of -100 volts, ID25 -90 amps, RDS(on) <=25mΩ. The IXTT96N20P is an N-channel MOSFET with a rise time of 160 ns, and maximum values: power dissipation 600 W, drain-source voltage 200 V, gate-source voltage 20 V, gate-threshold voltage 5 V, drain current 96 A, and drain-source on-state resistance 24 mOhms. In the current sense stage the C14 is a capacitor to filter out voltage spikes and let the DC pass through. A value of 100 nF is generally used to smooth out the higher frequencies and transient edges. The instrumentation amplifier (AD8221ARZ) is a classic three op-amp design, with 80 dB common mode rejection ratio (CMRR) up to 10 kHz (unity gain) and a 2 volt per microsecond slew rate, with 8 nV/root Hz noise at 1 kHz.

The current slew-rate is a function of the maximum voltage (15 volts), maximum current used (10 Amps – although the full range of the amplifier is up to 30 Amps), and coil inductance and resistance, and was measured to be approximately 30 T/m/s. Each gradient amplifier produces an output gain of 3.3 Amps per 1 volt input from the +/- 10 volt 16-bit digital-to-analogue (DAC) gradient drivers of the MR console. Figure 4 shows the gain as a function of input voltage, indicating a linearity of the output current as a function of input voltage of less than 1%. First order RC filters ($f_c$ = 7kHz ) were placed on the gradient lines to reduce RF noise introduced by the gradient amplifiers and gradient lines entering the Faraday cage.

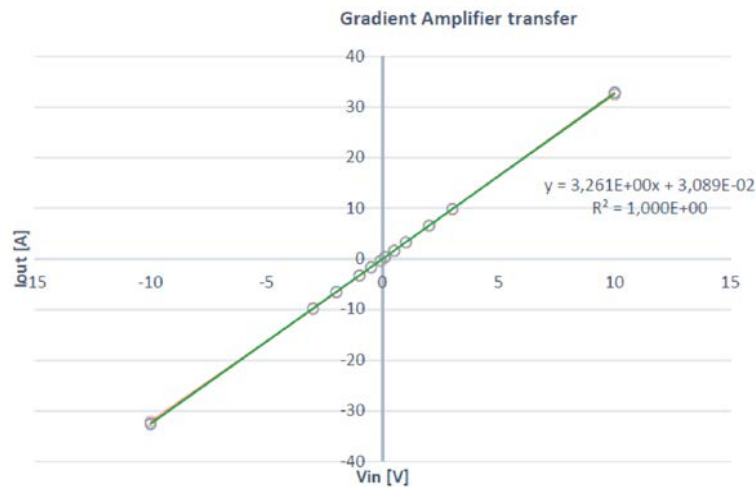

*Figure 4. Plot of the linearity of the output current of the gradient amplifier as a function of the voltage supplied by the digital-to-analogue (DAC) boards of the MR spectrometer.*

A Magritek Kea2 spectrometer (Aachen, Germany) was used to generate gradient wave forms and RF pulses as well as digitising the generated echoes. The RF pulses generated by the spectrometer were amplified by a custom built 1 kW RF amplifier (schematics are provided in supporting information figure 1).

*Phantom studies.*

Images were acquired of a phantom consisting of forty-five 35 mm long, 8 mm diameter tubes filled with water arranged on 3D printed a 7 × 7 rectangular grid (4 corner holes empty) with the centre of the tubes spaced 17 mm apart. Data was acquired using a 2D spin-echo projection sequence with the following scan parameters: field-of-view (FoV) 256 x 256 mm, 128 x 128 data points, TR/TE: 6000 ms/100 ms, pulse duration: 100 μs. Images were acquired with the main axis of the phantom in the +y (vertical) and +x (horizontal) axes, both perpendicular to the $B_0$ field.

*In-vivo imaging protocols*

All images were acquired from healthy volunteers giving informed consent according to the policies of the local ethics committee.

Brain imaging: all data were acquired with a bandwidth of 20 kHz, a 90° pulse length of 100 μs with a power of 47 dBm and a 180° pulse length of 100 μs with a power of 50 dBm. Phase cycling of successive 180° pulses in the TSE train followed the regular Carr Purcell Meiboom Gill (CPMG) modulation. K-space acquisition was restricted to an elliptical coverage in the two phase encoding directions in order to reduce scan time. All data were filtered using a sine-bell squared filter. $B_0$ homogeneity over the brain, measured as the Full-Width-at-Half-Maximum of the Fourier transform of the FID signal, was around 600 Hz.

Since the aim of the work is to map areas of CSF, most easily seen as bright areas on a darker background, we used a $T_2$-weighted sequence with long effective echo time. Based on a series of low-resolution inversion-recovery images with inversion time (TI) varying between 100 and 2800 ms, the $T_1$ of CSF was estimated to be ~1750 ms, with the $T_2$ value assumed to be only slightly lower since it is well-known that at very low field $T_1 \approx T_2$ for mobile liquids. Both the $T_1$ and $T_2$ relaxation times of brain tissue are much shorter, being reported on the order of 500 and 250 ms, respectively [36]. Three types of images were run: $T_2$-weighted for bright CSF, $T_1$-weighted for a rapid anatomical image (CSF dark), and an inversion-recovery sequence which produces good contrast but measurable signal intensities from both CSF and brain tissue. The spatial resolution was chosen to fulfil the relatively coarse requirements for pediatric imaging (hydrocephalus), within a very short imaging time of 2 minutes per 3D data set.

The sequence parameters for the sequences are as follows: $T_1$-weighted imaging: FoV: 192 × 192 × 256 mm, acquisition matrix: 48 × 48 × 64, TR/TE: 500 ms/10 ms, echo train length (ETL): 6, acquisition time: 2 minutes 23 seconds. K-space data were acquired on a centre-out trajectory. $T_2$ weighted: FoV: 160 × 192 × 256 mm, acquisition matrix: 40 × 48 × 64, TR/TE: 3000 ms/20 ms, ETL: 40, effective TE: 410 ms, , acquisition time: 1 minutes 48 seconds. K-space data were acquired on a low-high linear trajectory. Inversion recovery: FoV: 160 × 192 × 256 mm, acquisition matrix: 40 × 48 × 64, TI/TR/TE: 100 ms/3000 ms/20 ms, ETL: 40, acquisition bandwidth: 20 kHz, acquisition time: 1 minutes 48 seconds. K-space data were acquired using a centre-out trajectory.

In addition to these relatively low-resolution images, we also acquired a higher resolution image to demonstrate that the SNR is sufficient to acquire images at this resolution. FoV: 256 × 256 × 256 mm, acquisition matrix: 64 × 128 × 128, TR/TE: 500 ms/20 ms, ETL: 4, acquisition bandwidth: 20 kHz, acquisition time: 13 minutes 2 seconds. K-space data were acquired on a centre-out trajectory.

Knee imaging: for an illustrative case in which higher spatial resolution is desirable, and imaging times can be increased somewhat, we imaged the knee of a healthy volunteer. Relaxation data for muscle, fat and cartilage were estimated from previously published formulae and measurements [37]. The estimated $T_1$ values were 200, 140 and 70 ms, and $T_2$ values 47, 84 and 20 ms, respectively. In order to obtain signal from all tissues (with no attempt to optimize contrast between tissues), data were acquired using a 3D T1-weighted turbo spin echo sequence with the following parameters: FoV: 256 × 256 × 256 mm, acquisition matrix: 128 × 128 × 128, TR/TE: 115 ms/15 ms, ETL: 3, acquisition bandwidth: 20 kHz, RF pulse duration: 100 μs. K-space data were acquired on a centre-out trajectory and filtered using a sine-bell squared filter. $B_0$ homogeneity over the knee, measured as the Full-Width-at-Half-Maximum of the Fourier transform of the FID signal, was around 400 Hz.

*Gradient non-linearity correction*

Fourier-based image encoding assumes a linear relationship between space and gradient magnitude. Due to the relatively low length-to-diameter ratio (~1:1) possible for our gradient coils there are significant image distortions arising from gradient non-linearity, particularly along the x-axis (along the bore of the magnet). A basic distortion correction is implemented by replacing the standard inverse FFT by:

$$\rho(x,y,z) = \sum_x \sum_y \sum_z s(k_x, k_y, k_z) e^{i2\pi k_x G_x(x,y,z)} e^{i2\pi k_y G_y(x,y,z)} e^{i2\pi (G_z(x,y,z) + \Delta B_0(x,y,z))t}$$

where ρ is the reconstructed proton density, $G_x$, $G_y$ and $G_z$ are simulated gradient field maps (from Biot-Savart calculations using the wire geometries), and t the readout time. This approach to gradient non-linearity correction is analogous to that used in conjugate phase reconstruction. The algorithm was implemented in using Python 3.7 and Numpy 1.17: Numpy's optimized einsum function was used to reduce memory consumption and reconstruction time. Reconstruction time for a 128 x 128 x 128 voxel dataset was 43 seconds on an Intel Core i7-6700 quad-core CPU with 32 GB RAM. The algorithm was also adapted to run on a GPU using the Cupy python library. Reconstruction time for a 128 x 128 x 128 voxel dataset when executed on a Nvidia GTX 1660 super with 6GB on-board RAM was less than 1 second.

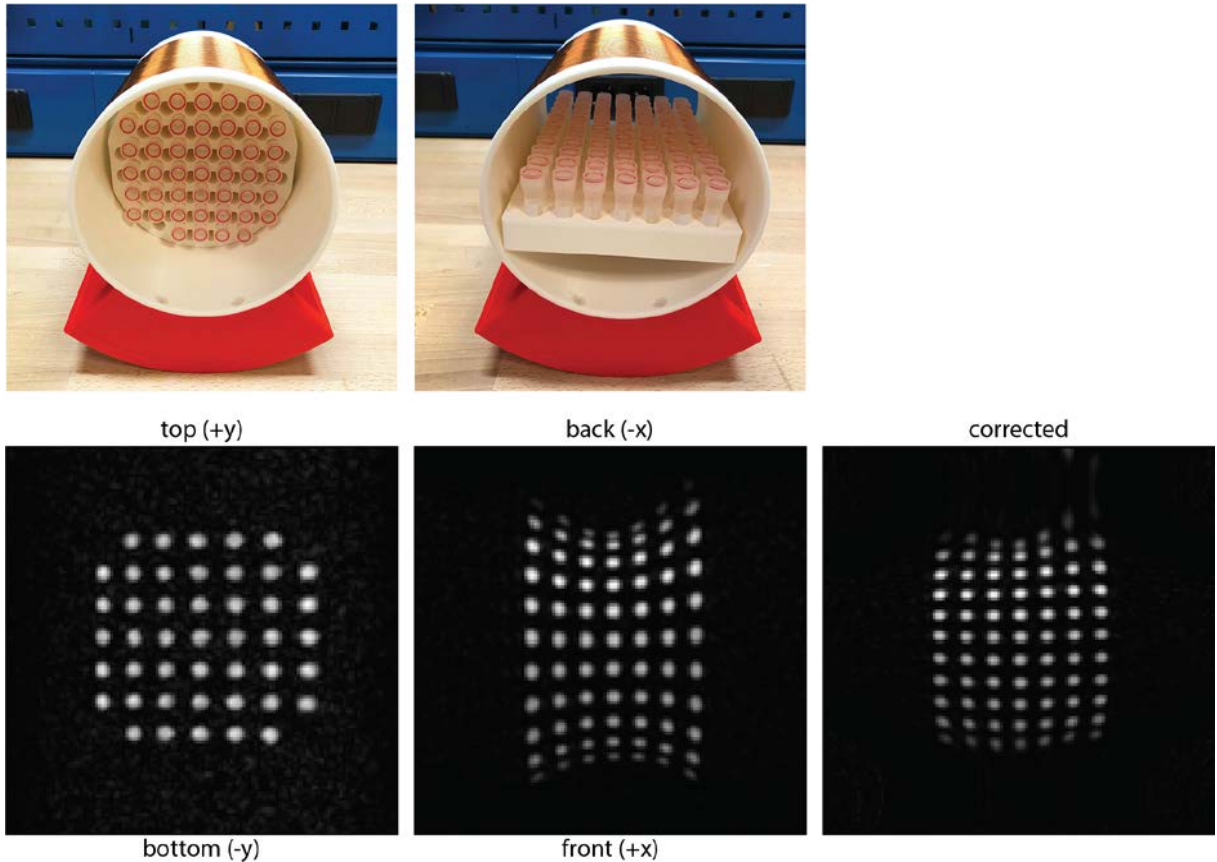

*Figure 5. Images acquired of a phantom of tubes (8 mm diameter, 30 mm long) each placed 17 mm apart on a rectangular grid. Minimal distortions are seen in the central transverse plane, more significant distortions are present in the central coronal plane and are primarily caused by non-linearities in the x gradient field.*

**Results**

Figure 5 shows 2D projection images phantom images, which illustrates the degree of geometric distortion particularly along the x-axis due to the non-linearity, and eventual compression point of the x-gradient coil. Also shown is the result of the correction algorithm which improves the reconstruction along this axis quite considerably.

Figure 6 shows a plot of the main field drift during experiments, measured from the resonance frequency of the proton spectrum acquired from the whole sample. For the phantom it is clear that the gradients produce negligible heating (and therefore field drift) in the experiment, and so correction does not need to be applied. For the in vivo data, the heating introduced by the body is much greater. The rate of drift is approximately 800 Hz per hour, corresponding to 160 Hz over our in vivo imaging time of twelve minutes.

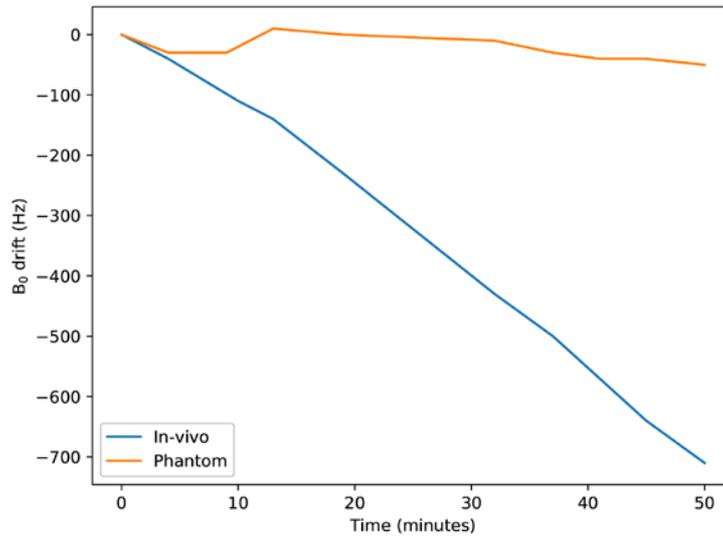

*Figure 6. Measurement of the $B_0$ drift measured spectroscopically during the phantom and in vivo experiments. Data were acquired over the entire sample within the active volume of the RF coil.*

Figure 7 shows in-vivo images of the brain of a healthy volunteers, with a slice through the centre of the ventricles shown since this is the area of most interest with respect to applications in hydrocephalus.

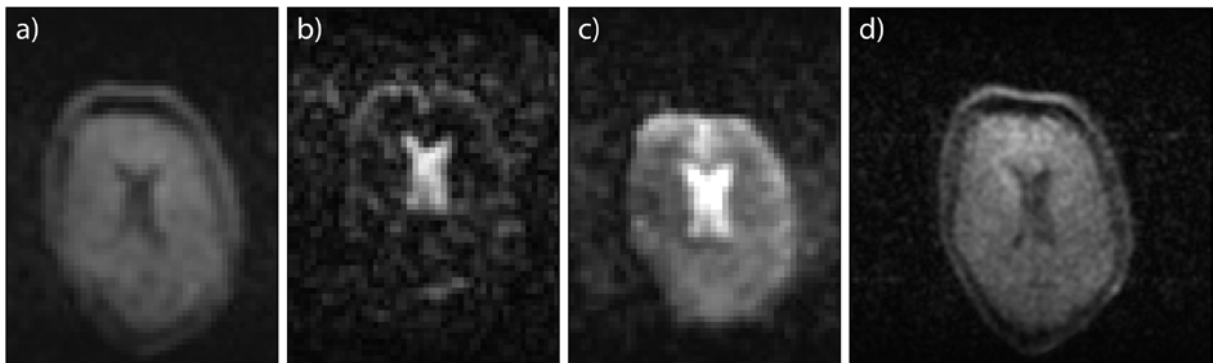

*Figure 7. Images acquired with different weighting at a spatial resolution of 4 × 4 × 4 mm. (a) T1- weighted (2.5 mins data acquisition time), (b) T2-weighted (2 mins) and (c) inversion-recovery TSE sequence (2 mins). (d) Higher resolution image (2 × 2 × 4 mm) from a different volunteer (13 minutes).*

Figure 8 shows in-vivo images of the knee acquired from a healthy volunteer. A single three-dimensional data set was acquired, and the images shown are three central and 20 mm offset planes through the data set.

Figure 9 shows the results of the gradient non-linearity correction, which shows improvement in the centre of the field-of-view, but as expected is unable to deal with the severe non-linearities close to the gradient null and reversal points at either ends of the gradient coil.

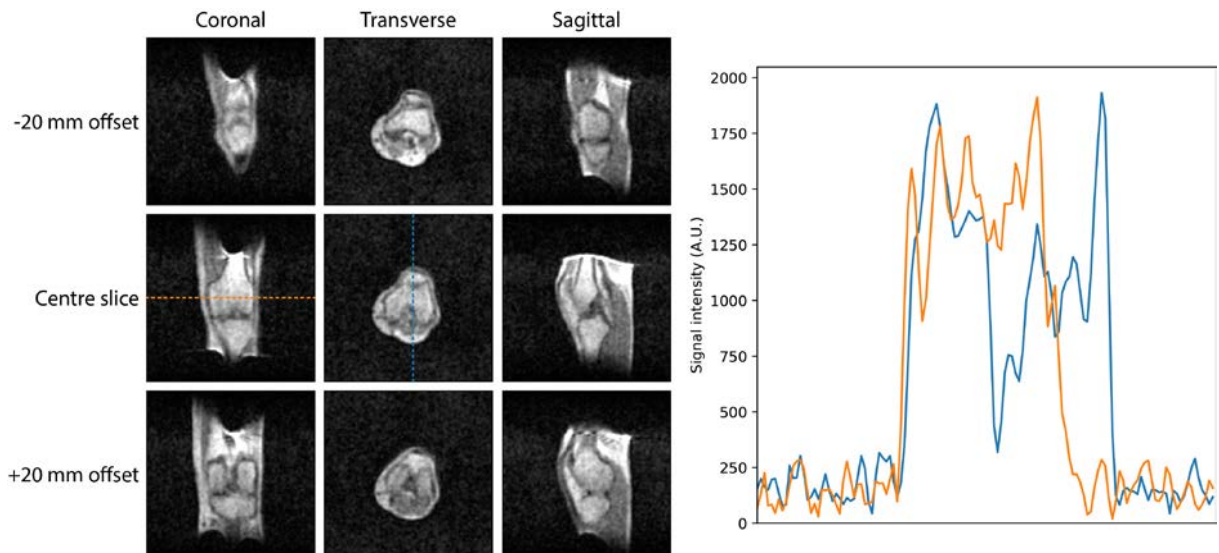

*Figure 8. (left) coronal, transverse and sagittal (reformatted from a single 3D acquisition) knee images using a 3D TSE sequence with the following parameters: FoV: 256 × 256 × 256 mm, Acq. Matrix: 128 × 128 × 128, TE/TR = 10 ms/ 130 ms, acq. time: 11 minutes 50 seconds. No gradient distortion correction has been applied to the images. (right) Projections taken through two of the central axes of the images.*

**Discussion**

We have designed and constructed a low-cost portable low-field MRI system operating at 50 mT for in vivo imaging. We would characterize the novelties of the problem of compactness, low cost and sustainability as the following:

1. Being able to design a permanent magnet array with readily available commercial small magnets which are inexpensive, i.e. do not require very fine tolerances,
2. Being able to form the housing for such an array with readily available and inexpensive plastics, which can be readily machined with acceptable tolerances,
3. Potentially being able to repair the magnet very simply by replacing individual magnets or rings of magnets should damage occur,
4. Having each of the electronics components which can run of standard power outlets, and ultimately could be battery/solar/diesel engine powered,
5. Designing gradient and RF amplifiers which operate for the specific tasks required (gradient slew times, operating frequency) rather than being multi-purpose much more expensive and heavy units.

In terms of power consumption and requirements, the radiofrequency amplifier runs off standard 110/240 volt mains supply and has to supply ~500 Watts at a duty cycle of <1%. The gradient amplifier also runs off a standard 110/240 volt mains supply and has to provide 300 VA at a duty cycle of ~20%. In terms of heating, the RF amplifier and gradient amplifier both contain heat sinks and have simple air fan cooling. A thermal camera indicated that the maximum component temperature reached in the gradient amplifier was ~60°C.

In vivo images of the brain and the knee have been acquired. For the brain scans, our ultimate aim is to image pediatric hydrocephalus in resource-limited settings. As discussed by Obungoloch et al. [38] the required spatial resolution is not high by conventional standards, and so we determined how rapidly it is possible to acquire images with sufficient spatial resolution. These were acquired within a couple of minutes, with sufficient contrast between brain tissue (WM/GM) and CSF to be able to determine areas of fluid accumulation.

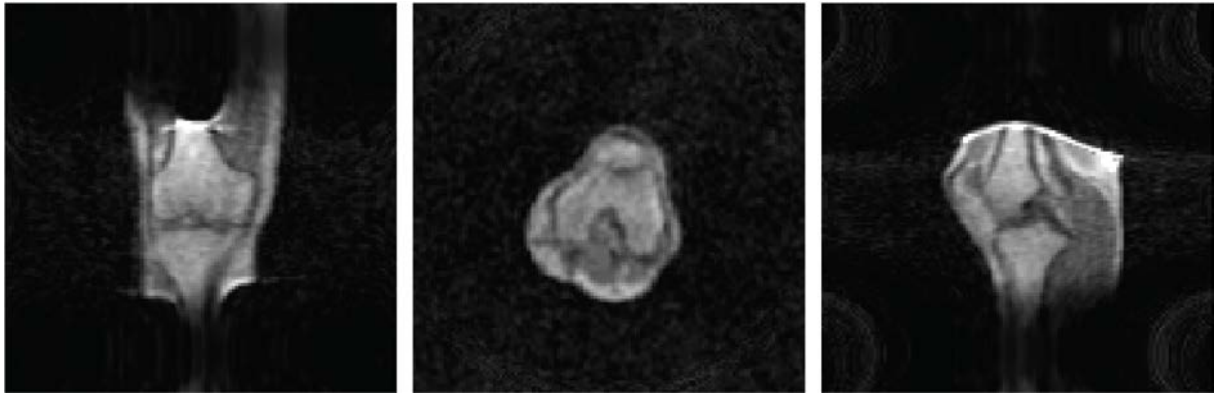

Figure 9. Central slices after applying the gradient correction algorithm. The remaining distortions in the image are located in areas of very strong gradient non-linearity and reversal points which the algorithm is unable to compensate for.

This type of system can also be used for applications in which the imaging time can be increased, and could be used as a point-of-care device. As an example, we showed that we can acquire in vivo images of the human knee with a nominal 2 × 2 × 2 mm resolution with a SNR of ~20:1 within twelve minutes.

In order to estimate the true resolution, there are several components in addition to digital resolution which must be considered. These include frequency drifts due to temperature changes (in vivo and gradient induced), the degree of $B_0$ inhomogeneity over the imaging region-of-interest with respect to the strength of the frequency encoding gradient, and $T_2$ decay during the TSE train. Taking each in turn, temperature induced $B_0$ drift causes by heat radiated from the body (the thermal sensitivity, due to the temperature sensitivity of the magnet remanence, is approximately 2.5 kHz per degree Celsius) were measured for the knee scans to be approximately 160 Hz over the twelve minute imaging session, which corresponds to a full-width-half-maximum (FWHM) of the point spread function (PSF) in the frequency encoding dimensions of around 1.9 mm for the 240 mm FOV and 20 kHz bandwidth used. However, these are very easily corrected by applying a corresponding phase shift to the k-space data. The NMR-measured inhomogeneity over the knee corresponded to a FWHM of ~400 Hz: imaging with a pixel bandwidth of 300 Hz/pixel gives a PSF of 1.3 pixels, or 2.6 mm in the frequency-encoding dimension. The echo train length of 3 echoes with an effective echo time of 15 ms and estimated $T_2$ of muscle of ~200 ms mean that the point spread function due to $T_2$ decay during the echo train is sub-pixel [39] in the corresponding phase encoding direction. Combining the effects of sampling and linewidth gives an estimated resolution of 3.3 mm in the frequency encoding, and ~2 mm in each of the phase encoding directions. Of course the former resolution could be reduced by using a higher readout bandwidth per pixel, at the cost of reduced SNR, using the gradients to perform first-order $B_0$ shimming and including knowledge of the $B_0$ field distribution in the image reconstruction. Similar calculations can be applied to the brain data, but since these are acquired at a coarser digital resolution of 4 × 4 × 4 mm, there is a correspondingly smaller effect. Several methods can also potentially be used to reduce the total imaging time: simple half Fourier, compressed sensing and parallel imaging using multiple receivers.

Noise measurements performed with each of the components of the system present and absent indicate that there are some improvements still to be made in terms of SNR. With a 50 Ω terminated input to the Magritek spectrometer the noise level is 0.18 microvolts for a 100 kHz bandwidth (the rms noise level is compared to an internal 1 microvolt reference signal produced by the spectrometer), corresponding to 0.7 nV/square root Hertz, which is within experimental error of the theoretical limit of 0.9 nV/square root Hertz. This level remains unchanged when the 50 Ohm load is located at the end of a long coaxial cable, which is passed through the BNC interface to the inside of the Faraday cage. When the impedance matched RF coil is connected, then the noise level is unchanged. Connecting the RF amplifier introduces no additional noise, but when the gradient amplifier is turned on the noise level increases to 0.33 microvolts. This indicates that some degree of coupling exists between the gradient coils and RF coil. The first order gradient filter reduces this by approximately 10%, but there is still room for improvement in reducing the noise level. We also note that there are a series of harmonic spikes which occur every 200 kHz, arising from the gradient amplifier: however, given the small imaging bandwidth used this is not a problem with the current setup.

We also anticipate that the Halbach magnet geometry can be expanded for future applications, moving away from the very simple cylindrical symmetrical structure illustrated here, for example hemispherical [40] or linear geometries [41] that might be appropriate for breast and adult head, and spine imaging, respectively.

**Acknowledgements**. This work was supported by Horizon 2020 European Research Grant FET-OPEN 737180 Histo MRI, Horizon 2020 ERC Advanced NOMA-MRI 670629, Simon Stevin Meester Prize and NWO WOTRO Joint SDG Research Programme W 07.303.101. We are grateful to Drs. Martin van Gijzen and Rob Remis at the TU Delft for collaborative discussions. We would also like to thank Dr. Lukas Winter for advice on RF amplifier construction as well as the design of the robot used for mapping the static magnetic field.

**Figure captions:**

Figure 1. A) Photograph of the magnet setup. The shims and gradients are integrated in to the bore of the magnet. The entire setup is placed inside an aluminium Faraday cage to reduce environmental noise in the setup. B) A conductive cloth is placed over the subject to reduce the noise coupled in to the system through the body.

Figure 2. Three imaging gradients are designed using the Target Field Method, adapted for the Halbach configuration with the $B_0$ field oriented across the magnet bore (left schematic): the line colour indicates the direction of current flow. The gradients were constructed using copper wires placed into a 3 mm-thick 3D printed structure.

Figure 3. Schematic of the three modules used in each of the three gradient amplifiers channels.

Figure 4. Plot of the linearity of the output current of the gradient amplifier as a function of the voltage supplied by the digital-to-analogue (DAC) boards of the MR spectrometer.

Figure 5. Images acquired of a phantom of tubes (8 mm diameter, 30 mm long) each placed 17 mm apart on a rectangular grid. Minimal distortions are seen in the central transverse plane, more significant distortions are present in the central coronal plane and are primarily caused by non-linearities in the x gradient field.

Figure 6. Measurement of the $B_0$ drift measured spectroscopically during the phantom and in vivo experiments. Data were acquired over the entire sample within the active volume of the RF coil.

Figure 7. Images acquired with different weighting at a spatial resolution of 4 × 4 × 4 mm. (a) $T_1$-weighted (2.5 mins data acquisition time), (b) $T_2$-weighted (2 mins) and (c) inversion-recovery TSE sequence (2 mins). (d) Higher resolution image (2 × 2 × 4 mm) from a different volunteer (13 minutes).

Figure 8. (left) coronal, transverse and sagittal (reformatted from a single 3D acquisition) knee images using a 3D TSE sequence with the following parameters: FoV: 256 × 256 × 256 mm, Acq. Matrix: 128 × 128 × 128, TE/TR = 10 ms/ 130 ms, acq. time: 11 minutes 50 seconds. No gradient distortion correction has been applied to the images. (right) Projections taken through two of the central axes of the images.

Figure 9. Central slices after applying the gradient correction algorithm. The remaining distortions in the image are located in areas of very strong gradient non-linearity and reversal points which the algorithm is unable to compensate for.

**Supplementary Information**
*Please contact the author for the supplementary information.*

Supporting Information Figure 1. Schematic of a custom built 1KW RF amplifier.

Supporting Information Figure 2. A 3D model of the X gradient coil wire pattern designed using the target field method described in [31].

Supporting Information Figure 3. A 3D model of the Y gradient coil wire pattern designed using the target field method described in [31].

Supporting Information Figure 4. A 3D model of the Z gradient coil wire pattern designed using the target field method described in [31].